\documentclass[reprint,amsmath,amssymb,aps,prl]{revtex4-2}

\usepackage{graphicx}
\usepackage{dcolumn}
\usepackage{bm}

\begin{document}


\title{Polarized Fock States for Polariton Photochemistry}

\author{Arkajit Mandal}
\email{amandal4@ur.rochester.edu}
\affiliation{Department of Chemistry, University of Rochester}
\author{Sebastian Montillo Vega}%
\affiliation{Department of Chemistry, University of Rochester}
\author{Pengfei Huo}
\email{phuo@ur.rochester.edu}
\affiliation{Department of Chemistry, University of Rochester}


\date{\today}

\begin{abstract}
We use the polarized Fock states to describe the coupled molecule-cavity hybrid system in quantum electrodynamics. The molecular permanent dipoles polarize the photon field by displacing its vector potential, leading to non-orthogonality between the Fock states of two different polarized photon fields. These polarized Fock states allow an intuitive understanding of several new phenomena that go beyond the prediction of the quantum Rabi model, and at the same time, offer numerical convenience to converge the results. We further exploit this non-orthogonality to generate multiple photons from a single electronic excitation (downconversion) and control the photochemical reactivity.
\end{abstract}
\maketitle
Coupling molecular systems to an optical cavity can significantly alter their potential energy landscape~\cite{Hutchison12,Ebbesen16,Kowalewski2017,Thomas2019}, and enable new chemical reactivities beyond the existing paradigms of chemistry. Existing theoretical framework for describing such molecule-cavity hybrid system is based upon adiabatic electronic states for the molecular subsystem and the Fock states of the vacuum field for the quantized radiation mode inside the cavity~\cite{Kowalewski2016,KowalewskiJCP2016,Herrera16,Galego2016,Flick2017PNAS,Feist2018,csehi2019ultrafast,Nitzan2019,Juan2020, Mandal2019JPCL,Galego2019,Vendrell2018,Bing2020,DU2019,Szidarovszky18}. While the adiabatic-Fock state is commonly used for describing matter-cavity interactions~\cite{Jaynes,Tavis}, it might not be the most convenient representation for describing strong interactions between the matter and the cavity~\cite{Bennett2016,Christian2018,Irish2005,Irish2007,Brumer2011,Jia2012}. In this work, we present a new representation based on the idea of the polarized Fock states, which allows one to intuitively understand new phenomena that go beyond the prediction of the quantum Rabi model.

We start by considering the Pauli-Fierz (PF) non-relativistic QED Hamiltonian~\cite{Rubio2018JPB,Rubio2018PF,Vendrell2018} to describe the light-matter interaction. The PF Hamiltonian can be rigorously derived~\cite{Christian2018,Nitzan2019,Mandal2020} (see SI for details) by applying the Power-Zienau-Woolley Gauge transformation~\cite{PZW,Cohen-Tannoudji} and a unitary phase transformation\cite{Mandal2020} on the minimal-coupling Hamiltonian in the Coulomb gauge ({\it i.e.} the ``$\mathrm{p\cdot A}$" Hamiltonian) under the long-wavelength limit. For a molecule coupled to a single photon mode inside an optical cavity, the PF QED Hamiltonian is
\begin{align}\label{eqn:PF}
\hat{H} &= \hat{H}_\mathrm{M}+{1\over2}\hat{p}^2 + {1\over2}\omega_\mathrm{c}^2 \Big(\hat{q} + \sqrt{\frac{2}{\hbar\omega_\mathrm{c}^3}}{\boldsymbol{\chi} \cdot \hat{\boldsymbol\mu}} \Big)^2 \\
&= \hat{H}_\mathrm{M}+\big(\hat{a}^{\dagger}\hat{a} + {1\over2}\big)\hbar \omega_\mathrm{c} + {\boldsymbol\chi }\cdot
\hat{\boldsymbol\mu}(\hat{a}^{\dagger} + \hat{a}) + {({\boldsymbol\chi}\cdot\hat{\boldsymbol\mu})^2\over \hbar \omega_\mathrm{c}}.\nonumber 
\end{align}
In the last line of the above equation, $\hat{H}_\mathrm{M}$ represents the molecular Hamiltonian, the second term $\hat{H}_\mathrm{P}=\big(\hat{a}^{\dagger}\hat{a} + {1\over2}\big)\hbar \omega_\mathrm{c}$ represents the Hamiltonian of the vacuum photon field inside the cavity with the frequency $\omega_\mathrm{c}$, the third term describes the light-matter interaction in the electric-dipole ``$\mathrm{d\cdot E}$" form ~\cite{Cohen-Tannoudji}, with $\boldsymbol{\chi}= \sqrt{\hbar\omega_\mathrm{c}\over 2 \varepsilon_0 \mathcal{V}} \hat{\bf e}\equiv \chi \hat{\bf e}$ characterizing the light-matter coupling vector oriented in the direction of polarization unit vector $\hat{\bf e}$, $\mathcal{V}$ as the quantization volume for the photon field, and $\varepsilon_0$ as the permittivity inside the cavity. The last term is the dipole self-energy (DSE), which describes how the polarization of the matter acts back on the photon field~\cite{Flick2017PNAS}.
Further, $\hat{a}^{\dagger}$ and $\hat{a}$ are the photon creation and annihilation operator, $\hat{q} = \sqrt{\hbar/2\omega_\mathrm{c}}(\hat{a}^{\dagger} + \hat{a})$ and $\hat{p} = i\sqrt{\hbar\omega_\mathrm{c}/2}( \hat{a}^{\dagger} - \hat{a})$ are the photonic coordinate and momentum operator, respectively, and $\hat{\boldsymbol\mu}$ is the molecular dipole operator (for both electrons and nuclei). Throughout this study, we assume that $\hat{\bf e}$ align with the direction of $\hat{\boldsymbol\mu}$. The matter Hamiltonian is expressed as
\begin{equation}
\hat{H}_\mathrm{M}=\hat{T}_{R}+\hat{H}_\mathrm{el}(R,r), 
\end{equation}
where $\hat{T}_{R}=\hat{P}^2/2M=-\hbar^2 \nabla^2_{R}/2M$ is the nuclear kinetic energy operator,$\hat{H}_\mathrm{el}(R,r)=\hat{T}_{r}+\hat{V}_\mathrm{c}(R,r)$ is the electronic Hamiltonian, with the electronic kinetic energy $\hat{T}_{r}$ and Coulomb potential $\hat{V}_\mathrm{c}(R,r)$ among electrons and nuclei.

The polaritonic Hamiltonian is defined $\hat{H}_\mathrm{pl} \equiv \hat{H} - \hat{T}_{R}$, and the polariton surface $E_j(R)$ and polariton state $|\Phi_j (R)\rangle$ as the eigenvalue and eigenstate of $\hat{H}_\mathrm{pl}$ as
\begin{align}\label{eqn:Polariton}
\hat{H}_\mathrm{pl} |\Phi_j (R)\rangle \equiv (\hat{H} - \hat{T}_{R}) |\Phi_j (R)\rangle= E_j(R) |\Phi_j (R)\rangle. 
\end{align}
The commonly used basis to solve the above eigenequation is the adiabatic-Fock basis $\{|g\rangle \otimes |n\rangle,|e\rangle \otimes |n\rangle\}$, with eigenstates of the electronic Hamiltonian $\hat{H}_\mathrm{el}$, {\it i.e.}, the adiabatic electronic states $\{|g (R)\rangle, |e (R)\rangle\}$ (here we consider two of them) for the matter part, and the Fock states of the radiation mode (vacuum photon field) $\{|n\rangle\}$, {\it i.e.}, the eigenstate of $(\hat{a}^{\dagger}\hat{a} + {1\over2})\hbar \omega_\mathrm{c}$. For an atom, $\hat{H}_\mathrm{M}=E_{g}|g\rangle\langle g|+E_{e}|e\rangle\langle e|$, and the transition dipole is ${\boldsymbol \mu}_{eg}=\langle e|\hat{\boldsymbol\mu}|g\rangle$. Note that the permanent dipoles are $\boldsymbol\mu_{ee}=\langle e|\hat{\boldsymbol\mu}|e\rangle=0$, ${\boldsymbol\mu}_{gg}=\langle g|\hat{\boldsymbol\mu}|g\rangle=0$. Thus, the dipole operator is expressed as $\hat{\boldsymbol\mu}= \boldsymbol \mu_{eg}(|e\rangle \langle g| + |g\rangle \langle e|)\equiv\mu_{eg}(\hat{\sigma}^{\dagger} + \hat{\sigma})$ by defining the creation operator $\hat{\sigma}^{\dagger}\equiv|e\rangle \langle g|$ and annihilation operator $\hat{\sigma}\equiv|g\rangle \langle e|$ of the electronic excitation. The atom-cavity PF Hamiltonian becomes 
\begin{equation}\label{eqn:atom-cavity}
\hat{H}=\hat{H}_\mathrm{M}+\hat{H}_\mathrm{P}+{\boldsymbol\chi }\cdot{\boldsymbol\mu}_{eg}(\hat{\sigma}^{\dagger} + \hat{\sigma})(\hat{a}^{\dagger} + \hat{a})+{({\boldsymbol\chi}\cdot{\boldsymbol\mu}_{eg})^2\over \hbar \omega_\mathrm{c}}.
\end{equation}
Dropping the DSE (the last term) from Eqn.~\ref{eqn:atom-cavity} leads to the Rabi Model $\hat{H}_\mathrm{Rabi}=\hat{H}_\mathrm{M}+\hat{H}_\mathrm{P}+{\boldsymbol\chi }\cdot{\boldsymbol\mu}_{eg}(\hat{\sigma}^{\dagger} + \hat{\sigma})(\hat{a}^{\dagger} + \hat{a})$. Dropping both the DSE and the counter-rotating terms leads to the well-known Jaynes-Cummings Model~\cite{Jaynes} $\hat{H}_\mathrm{JC}=\hat{H}_\mathrm{M}+\hat{H}_\mathrm{P}+{\boldsymbol\chi }\cdot{\boldsymbol\mu}_{eg}(\hat{\sigma}^{\dagger}\hat{a}+\hat{\sigma}\hat{a}^{\dagger})$.

For a molecular system, we have $\hat{H}_\mathrm{M}=\sum_{ij}{1\over 2 M}(\hat{P}\delta_{ij}-i\hbar{d}_{ij})^{2} |i\rangle\langle j|+ E_{g}(R) |g \rangle\langle g|+ E_{e}(R) |e\rangle\langle e|$, where $\{i,j\}\in\{g,e\}$, ${d}_{ij}=\langle i|\nabla|j\rangle$ is the non-adiabatic coupling. The dipole operator has the following expression
\begin{equation}\label{eqn:dipole}
\hat{\boldsymbol\mu} = {\boldsymbol\mu}_{gg} (R) |g\rangle \langle g| + {\boldsymbol \mu}_{ee} (R) |e\rangle \langle e| + {\boldsymbol \mu}_{eg}(R)\big(|e\rangle \langle g| + |g\rangle \langle e|\big),
\end{equation}
where the adiabatic permanent dipoles are not zero. We further express the dipole operator in its eigenstate representation as
\begin{equation}\label{eqn:dipole-dia}
\hat{\mu}= {\boldsymbol\mu}_\mathrm{I}(R) |\mathrm{I}\rangle \langle \mathrm{I}| + {\boldsymbol\mu}_\mathrm{C} (R) |\mathrm{C}\rangle \langle \mathrm{C}|=\sum_{\alpha} {\boldsymbol\mu}_{\alpha}(R)|\alpha\rangle \langle \alpha|.
\end{equation}
Here, the {\it eigenstates} of $\hat{\boldsymbol\mu}$ are denoted as the covalent state $|\mathrm{C}\rangle$ and the ionic states $|\mathrm{I}\rangle$ for a diatomic molecule, and $\alpha\in\{\mathrm{I}, \mathrm{C}\}$. For a two-state system, analytical results are available (see SI), and for multiple electronic states, one can directly diagonalizing the adiabatic dipole matrix to obtain them. Diagonalizing Eqn.~\ref{eqn:dipole} to obtain Eqn.~\ref{eqn:dipole-dia} is commonly referred to as the Mulliken-Hush diabatization~\cite{ Mulliken1952,Cave1996,Cave1997,Hush2007,Timothy2004}, where the $\{|\mathrm{C}\rangle, |\mathrm{I}\rangle\}$ are commonly used as approximate {\it diabatic} states that are defined based on their characters (covalent and ionic). In this work, we explicitly assume that $|\mathrm{I}\rangle$ and $|\mathrm{C}\rangle$ are strict diabatic states, hence $\langle \mathrm{C}|\nabla_R|\mathrm{I}\rangle=0$ (they are $R$-independent). This assumption simplifies our argument, but will not impact any conclusion we draw (see SI for details). 
Under the special case of the atomic cavity QED where ${\boldsymbol \mu}_{ee} = {\boldsymbol \mu}_{gg} = 0$, $\hat{\boldsymbol\mu}={\boldsymbol\mu}_{eg}|+\rangle\langle+|-{\boldsymbol\mu}_{eg}|-\rangle\langle -|$ and $|\pm\rangle=[|g\rangle\pm|e\rangle]/\sqrt{2}$ (the eigenstates of $\hat{\boldsymbol \mu}$) are referred to as the the qubit states~\cite{Irish2005,Irish2007}. 

With the eigenstate of $\hat{\boldsymbol\mu}$ (diabatic state),  the molecular Hamiltonian becomes $\hat{H}_\mathrm{M} = \hat{T}_{R} + \sum_{\alpha} V_{\alpha}(R) |{\alpha}\rangle \langle {\alpha}| + V_\mathrm{IC}(R) (|\mathrm{C}\rangle \langle \mathrm{I}| + |\mathrm{I}\rangle \langle \mathrm{C}|)$, where $V_\alpha(R)$ represents the diabatic potentials, $V_{\mathrm{IC}}(R)$ represents the diabatic coupling. The PF Hamiltonian in Eqn.~\ref{eqn:PF} under the $|\alpha\rangle$ is expressed as $\hat{H}=  \hat{H}_\mathrm{M}+{\hat{p}^2\over2}+\sum_{\alpha} [{\omega_\mathrm{c}^2\over2} (\hat{q}+ q^0_{\alpha}(R))^2]|\alpha\rangle \langle \alpha|$, where $q^0_{\alpha}(R)=\sqrt{2\over\hbar\omega_\mathrm{c}^3}{{\boldsymbol\chi}\cdot {\boldsymbol \mu_i}(R)}$. We notice that the photon field is described as displaced Harmonic oscillator that is centered around $-q^0_{\alpha}(R)$. This displacement can be viewed as a {\it polarization} of the photon field due to the presence of the molecule-cavity coupling, such that the photon field corresponds to a non-zero (hence polarized) vector potential, in contrast to the vacuum photon field. 

The {\it central idea} of this work stems from the {\it polarized Fock states} (PFS) defined as follows
\begin{align}\label{polfoc}
&\frac{1}{2}\big[\hat{p}^2+\omega_\mathrm{c}^2(\hat{q} + q^{0}_{\alpha}(R))^2\big]|n_{\alpha}(R)\rangle\\
&\equiv (\hat{b}^{\dagger}\hat{b}+\frac{1}{2})\hbar\omega_c|n_{\alpha}(R)\rangle=\big(n+\frac{1}{2}\big)\hbar \omega_c|n_{\alpha}(R)\rangle,\nonumber
\end{align}
where the PFS $|n_{\alpha}(R)\rangle \equiv |n_{\alpha}\rangle$ is the Fock state of a displaced Harmonic oscillator, with the displacement $-q^{0}_{\alpha}=-\sqrt{2\over\hbar\omega_\mathrm{c}^3}{{\boldsymbol \chi}\cdot {{\boldsymbol \mu}_\alpha}(R)}$ specific to the diabatic state $|\alpha\rangle$, and $n=0,1,2...,\infty$ is the quantum number.  Further, $\hat{b}^{\dagger}_{\alpha}=(\hat{q}'_\alpha+i\hat{p})/\sqrt{2}$ and $\hat{b}_{\alpha}=(\hat{q}'_\alpha-i\hat{p})/\sqrt{2}$ are the creation and annihilation operators of the PFS $|n_{\alpha}\rangle$, with the photon field momentum operator $\hat{p}$ and polarized photon field coordinate operator $\hat{q}'_\alpha = \hat{q}+q^{0}_{\alpha}(R)$. Compared to the vacuum's Fock state $|n\rangle$, these PFS depends on the diabatic state (or more generally, $\hat{\boldsymbol\mu}$'s eigenstate) of the molecule, and the position of the nuclei (through the $R$ dependence in ${\boldsymbol \mu}_{\alpha}(R)$). Due to the electronic state-dependent nature of the polarization (from the difference between ${\boldsymbol\mu}_{I}$ and ${\boldsymbol\mu}_\mathrm{C}$), the PFS associated with different electronic diabatic states becomes non-orthogonal, {\it i.e.},  $\langle n_\mathrm{I}|m_\mathrm{C}\rangle \neq \delta_{nm}$. The PFS is closely related to the polarized vacuum states~\cite{Christian2018}, with the key difference that $|n_{\alpha} (R)\rangle$ does not parametrically depend on the positions of electrons $\hat{r}$ (hence $\langle m_\mathrm{I}| \nabla_{r} | n_\mathrm{I}\rangle = 0$), while it does parametrically depends on the nuclear position $R$ such that $\langle m_\mathrm{I}| \nabla_R | n_\mathrm{I}\rangle \neq 0$. Under the special case of the atomic cavity QED, the PFS representation reduces to the qubit-shifted Fock basis used in the generalized rotating-wave approximation~\cite{Irish2005,Irish2007,Brumer2011}. 

With the PFS, we use the basis $|\alpha, n_{\alpha}\rangle\equiv|\alpha\rangle\otimes|n_{\alpha} (R)\rangle$ to evaluate the matrix elements of the PF Hamiltonian $\hat{H}=\hat{T}_{R}+\hat{H}_\mathrm{pl}$. These matrix elements can also be equivalently obtained (see SI for detail) by applying a polaron-type transformation~\cite{Nitzan2019}, $\hat{U}^{\dagger}_\mathrm{pol}\hat{H}\hat{U}_\mathrm{pol}$, where $\hat{U}_\mathrm{pol}=\exp[-{i\over\hbar}\hat{p} \sum_{\alpha}q_{\alpha}^{0}(R)|\alpha\rangle \langle \alpha|]$ is a photonic coordinate displacement operator. The polariton Hamiltonian $\hat{H}_\mathrm{pl}$ is expressed as
\begin{align}\label{eqn:Hpl}
\hat{H}_\mathrm{pl}&=\sum_{\alpha}\sum_{n}\big(V_{\alpha}(R) + (n + \frac{1}{2})\hbar \omega_\mathrm{c} \big) |\alpha, n_\alpha\rangle \langle \alpha, n_\alpha|  \\
&+ \sum_{n,m}\langle  m_\mathrm{C}|  n_\mathrm{I}\rangle V_\mathrm{IC}(R)\big(|\mathrm{I}, n_\mathrm{I}\rangle \langle \mathrm{C},m_\mathrm{C}| + |\mathrm{C},m_\mathrm{C}\rangle \langle \mathrm{I}, n_\mathrm{I}|\big). \nonumber
\end{align}
Note that there is a finite coupling between the ionic state with $n$ photons and the covalent state with $m$ photons through the $\langle m_\mathrm{C}| n_\mathrm{I} \rangle V_\mathrm{IC}(R)$ term, which is the diabatic electronic coupling $V_\mathrm{IC}(R)$ scaled by the overlap $\langle m_\mathrm{C}| n_\mathrm{I}\rangle$ of the PFS. Further, $\hat{T}_{R}$ in the $|\alpha, n_{\alpha}\rangle$ basis is given by
\begin{align}\label{eqn:NucT}
\hat{T}_{R}&= \sum_{\alpha,n,m}{1\over{2M}}\big({\hat{P}\delta_{n,m}- i\hbar\langle m_\alpha| \nabla_R |n_\alpha\rangle }\big)^2 |\alpha,m_\alpha\rangle \langle \alpha, n_\alpha|.
\end{align}
Note that there is no non-adiabatic couplings between states with different diabatic characters, since $\langle \mathrm{C},n_\mathrm{C}| \nabla_R |\mathrm{I}, m_\mathrm{I}\rangle=\langle n_\mathrm{C}| \nabla_R |m_\mathrm{I}\rangle \langle \mathrm{C}|\mathrm{I}\rangle=0$ (because we assume that $|\mathrm{I}\rangle$ and $|\mathrm{C}\rangle$ are strict diabatic basis), and they are orthogonal $\langle\mathrm{C}|\mathrm{I}\rangle=0$. The polaritonic non-adiabatic coupling can be  analytically evaluated (see details in SI) as $\langle m_\alpha| \nabla_R |n_\alpha\rangle= -{\boldsymbol{\chi}\over\hbar\omega_\mathrm{c}}\cdot{\nabla_R \boldsymbol{\mu}_{\alpha}(R) \langle m_\alpha | \hat{b}^{\dagger} - \hat {b}|n_\alpha\rangle}$. Thus, these terms couple off-resonant states that are separated by $\hbar\omega_\mathrm{c}$ through the $(\hat{b}^{\dagger} - \hat {b})$ term.  It plays a similar role as the vector potential in the light-matter Hamiltonian in the coulomb gauge. In fact, upon an unitary transformation $\hat{U}_\theta = \exp[-i{\pi}(\sum_{\alpha}\hat{b}_{\alpha}^\dagger \hat{b}_\alpha |\alpha\rangle\langle\alpha|]$, we can explicitly show (see SI) that $\hat{H}'=\hat{U}_\theta^\dagger \hat{H}_\mathrm{pl}\hat{U}_\theta$ adapts `p.A' form where {\it only} the nuclear momentum operator interacts with polarized vector potential $\hat{A}_\alpha={\boldsymbol\chi}\cdot \nabla_R {\boldsymbol\mu}_{\alpha}(R)(\hat{b}^{\dagger }_\alpha + \hat{b}_\alpha)/\omega_\mathrm{c}$. 

Combining Eqn.~\ref{eqn:Hpl} and Eqn.~\ref{eqn:NucT}, we have the full expression of the total Hamiltonian $\hat{H}=\hat{T}_{R}+\hat{H}_\mathrm{pl}$ under the $|\alpha, n_{\alpha}\rangle$ representation. From these detailed expressions in Eqn.~\ref{eqn:Hpl}-\ref{eqn:NucT}, one can clearly see that quantum transitions among $\{|\alpha, n_{\alpha}\rangle\}$ states are caused by scaled couplings $\langle m_\mathrm{C}|n_\mathrm{I}\rangle V_\mathrm{IC}(R)$ in $\hat{H}_\mathrm{pl}$, as well as non-adiabatic couplings $\langle m_\alpha| \nabla_R |n_\alpha\rangle$ in $\hat{T}_{R}$. For the range of the photon frequency used in this work, $\langle m_\alpha| \nabla_R |n_\alpha\rangle$ does not play any role in the dynamics (as numerically demonstrated in SI). However, their role should not be overlooked and will be explored in future. 

We conjecture that the basis $\{|\mathrm{I},n_\mathrm{I}\rangle, |\mathrm{C},m_\mathrm{C}\rangle\}$ is both computationally economic and conceptually intuitive than the conventional adiabatic-Fock states $|g,n\rangle,|e,m\rangle $. For numerical efficiency, we find that one only needs a few of $|\alpha, n_{\alpha}\rangle$ basis to converge the results of solving Eqn.~\ref{eqn:PF}, whereas one needs 2-20 times more vacuum's Fock states in the range of parameters used here. For quantum dynamics simulations, we have used both the $|\alpha, n\rangle$ and $|\alpha, n_{\alpha}\rangle$ basis to perform numerically exact simulations with the split-operator method~\cite{Tannor}, and discovered a similar numerical efficiency of the $|\alpha, n_{\alpha}\rangle$ basis. This is because that the vacuum's Fock states centers around $q=0$; one needs a lot of $|n\rangle$ to represent the hybrid system that involves light-matter interaction with a potential centered around $-q^{0}_{\alpha}$ (see Eqn.~\ref{polfoc}). Conceptually, it allows one to intuitively understand the existence of certain light induced avoid crossing which is not predicted by the Rabi model. To demonstrate these effects, we use a well parameterized diabatic model of the LiF molecule~\cite{Timothy2004} and investigate the molecule-cavity QED.
\begin{figure}
 \centering
  \begin{minipage}[h]{1.0\linewidth}
     \centering
     \includegraphics[width=\linewidth]{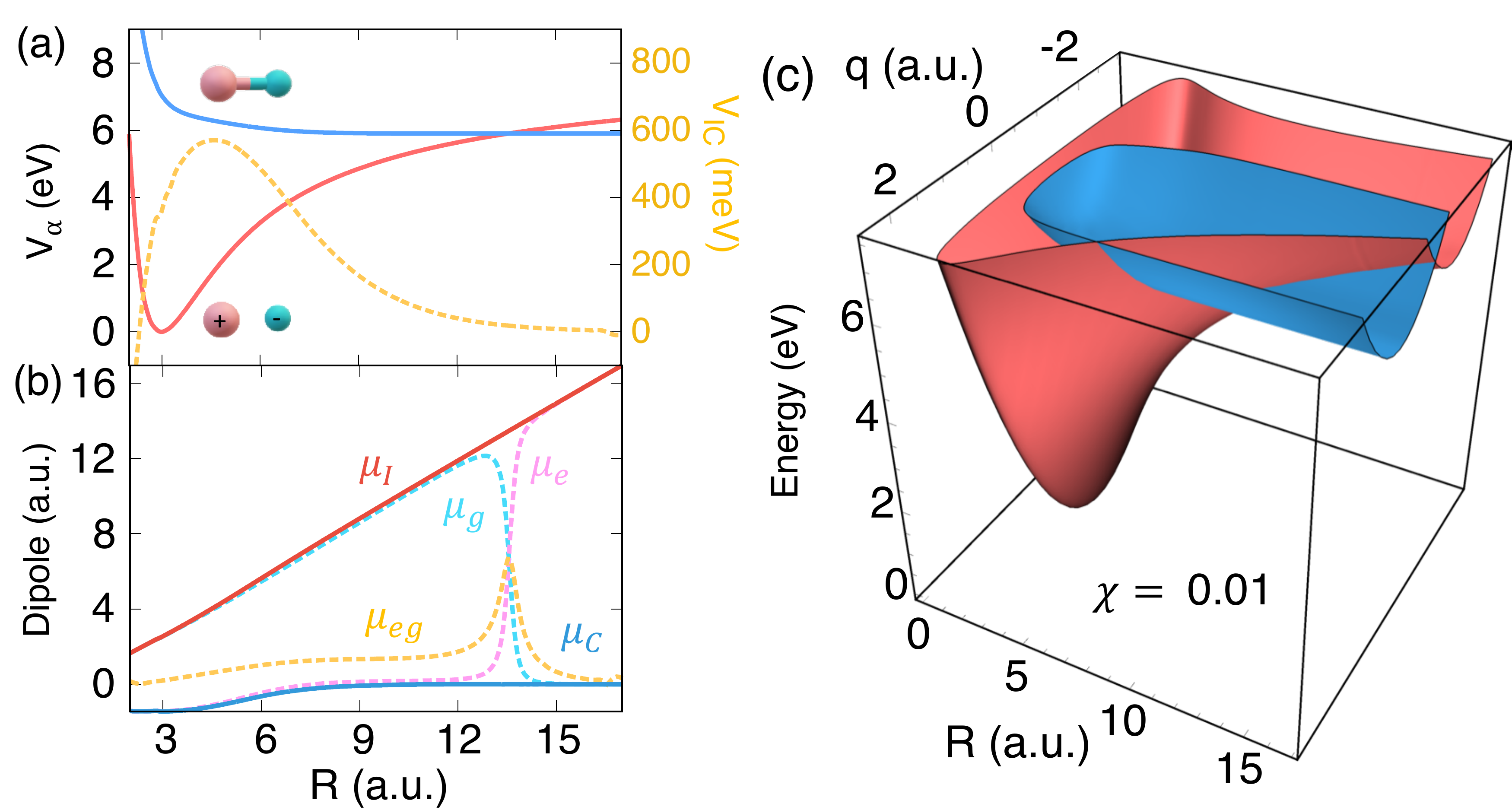}
       \end{minipage}%
   \caption{\small LiF model molecular system. (a) Diabatic potentials $V_{I}(R)$ (red) and $V_{C}(R)$ (blue), with diabatic coupling $V_\mathrm{IC}$ (gold line). (b) Matrix elements of $\hat{\mu}$ in the adiabatic representation (dashed curves) $\mu_{gg}$ (pink), $\mu_{ee}$ (cyan), and $\mu_{eg}$ (gold), as well as in the diabatic representation (solid lines) $\mu_\mathrm{I}$ (red) and $\mu_\mathrm{C}$ (blue). (c) Cavity diabatic potentials $V_{\alpha}(R)+{\omega_\mathrm{c}^2\over2}(\hat{q} + q^{0}_{\alpha}(R) )^2$ for the $|\alpha\rangle=|\mathrm{I}\rangle$ (blue) and $|\alpha\rangle=|\mathrm{C}\rangle$ (red) as a function of the nuclear coordinate $R$ and the photonic coordinate $q$.}
\label{Model}
\end{figure}

Fig.~\ref{Model}a presents the diabatic potentials energy surface $V_{\alpha} (R)$ of the $|\mathrm{I}\rangle$ (red) and $|\mathrm{C}\rangle$ state (blue), respectively. The crossing of these two {\it diabatic} curves occur at $R = R_0 \approx 13.5$ a.u., forming an avoided crossing between the adiabatic states $|g\rangle$ and $|e\rangle$ (not shown here). The diabatic coupling is $V_\mathrm{IC}(R)$ (gold line). Fig.~\ref{Model}b presents the matrix elements of $\hat{\boldsymbol\mu}$  in both the diabatic (solid lines) and the adiabatic (dashed lines) representations. The ionic permanent dipole (solid red) ${\boldsymbol\mu}_\mathrm{I}(R)$ increases linearly with $R$, while the covalent permanent dipole (solid blue) ${\boldsymbol \mu}_\mathrm{C}(R) \approx 0$, as one expects. The adiabatic states switch their characters around $R_0$, as a results, the adiabatic permanent dipole switches in that region, and $\mu_{eg}(R)$ peaks at $R_0$ as the two diabatic states couple strongly in their crossing region. Fig.~\ref{Model}c demonstrate the electronic state-dependent photon field polarization by visualizing $V_{\alpha}(R)+{\omega_\mathrm{c}^2\over2}(\hat{q} + q^{0}_{\alpha}(R))^2$. These diabatic surfaces are depicted as a function of $R$ and $q$, at $\chi = 0.01$ a.u. and $\hbar\omega_\mathrm{c} = 7.5$ eV. The surfaces are color-coded corresponding to their {\it diabatic} electronic characters $|\mathrm{I}\rangle$ (blue) and $|\mathrm{C}\rangle$ (red). The covalent diabatic surface along $q$ is not displaced because $\mu_\mathrm{C}(R)$ is nearly zero, and the ionic cavity diabatic surface is increasingly displaced along $q$ with an increasing $R$, because $\mu_\mathrm{I}(R)$ increases linearly along $R$. At a larger $R$, the extent of the photon field polarization is significantly different for the $|\mathrm{I}\rangle$ and the $|\mathrm{C}\rangle$ state. 

\begin{figure}
 \centering
  \begin{minipage}[h]{1.0\linewidth}
     \centering
     \includegraphics[width=\linewidth]{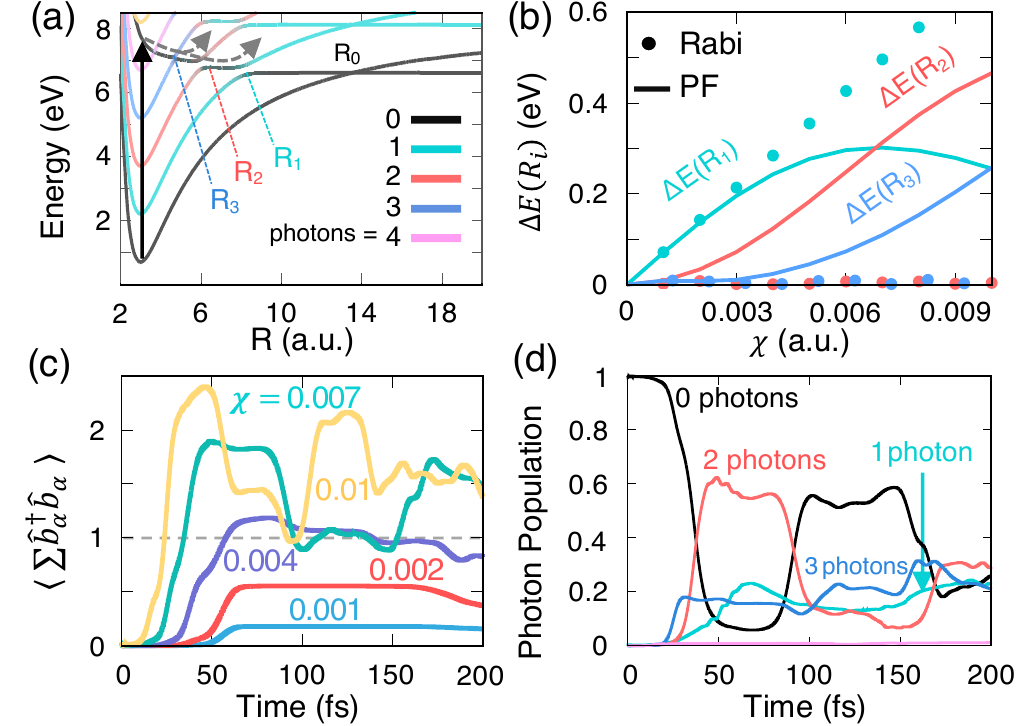}
       \end{minipage}%
   \caption{\small Using state-dependent polarization to perform downconversion. (a) Polaritonic potentials color coded according to the number of photons with four relevant avoided crossings labeled as $R_0-R_3$. The black solid vertical arrow indicates the initial photoexcitation, the dashed lines illustrate the dynamics of the hybrid system. (b) The energy-splitting at three avoided crossings as a function of $\chi$ computed from Pauli-Fierz and Rabi Hamiltonian. (c) Time-dependent expectation value of the number of photons at various  $\chi$. (d) Time-dependent photon populations at $\chi = 0.007$ a.u.}
\label{down}
\end{figure}

Fig.~\ref{down} demonstrates that the non-orthogonality between the PFS can be used to convert a single molecular excitation into multiple excitations in the cavity, {\it i.e}, a downconversion process. It can be seen from Eqn.~\ref{eqn:Hpl} that the covalent state with zero photons $|\mathrm{C},0_\mathrm{C}\rangle$ couples to $|\mathrm{I},n_\mathrm{I}\rangle$ through the coupling  $V_\mathrm{IC}(R)\langle  0_\mathrm{C}| n_\mathrm{I}\rangle$. Fig.~\ref{down}a presents the polaritonic potential energy surfaces with $\chi = 0.007$ a.u. and $\hbar\omega_\mathrm{c} = $1.5 eV. The polariton potential $E_j(R)$ associated with the state $|\Phi_j (R)\rangle$ (see Eqn.~\ref{eqn:Polariton}) is color coded according to the expectation value of the number of photons $\langle\hat{N}\rangle=\langle \Phi_j (R)|\sum_{\alpha}\hat{b}^{\dagger}_{\alpha} \hat{b}_{\alpha}|\Phi_j (R)\rangle$, where $\hat{b}^{\dagger}_{\alpha}$ and $\hat{b}_{\alpha}$ are the creation and annihilation operator of the PFS (see Eqn.~\ref{polfoc}), where $\alpha\in\{\mathrm{I,C}\}$. We emphasize that for a molecule-cavity hybrid system, $\langle\hat{N}\rangle=\langle\sum_{\alpha}\hat{b}^{\dagger}_{\alpha} \hat{b}_{\alpha}\rangle$ is the physically meaningful way to characterize the number of photon~\cite{Christian2020}, whereas using vacuum's number operator $\langle\hat{N}'\rangle=\langle\hat{a}^{\dagger} \hat{a}\rangle$ gives an un-physical measure~\cite{Christian2020}. 

In Fig.~\ref{down}a, several new light-induced avoided crossings (LIAC) at $R_1$, $R_2$, and $R_3$ are formed due to the light-matter interactions, in addition to the original electronic avoid crossing at $R_0$.

Fig.~\ref{down}b presents the energy-splitting $\Delta E (R_i)$ associated with three cavity-induced avoided crossings at $R_1$, $R_2$ and $R_3$ as a function of $\chi$. Here, we compare these energy-splittings computed from both the Rabi (filled circles) and the Pauli-Fierz Hamiltonian (solid lines). For the Rabi model in the adiabatic-Fock representation ($\{|g,n\rangle,|e,n\rangle\}$), one ignores the permanent dipole contribution (${\boldsymbol\mu}_{ee}$ and ${\boldsymbol\mu}_{gg}$), as well as all of the DSE terms. The Rabi model is widely used in recent molecular polariton chemistry investigations \cite{Kowalewski2016,Vendrell2018,Bennett2016}. While the Rabi model provides a reasonable description of $\Delta E (R_1)$ at a weak coupling, it fails to correctly describe $\Delta E (R_1)$ at a larger coupling strength, and failed to predict $\Delta E (R_2)$ and $\Delta E (R_3)$. This is because that these deviations are caused by permanent dipole moments ${\boldsymbol\mu}_{gg}$ and ${\boldsymbol\mu}_{ee}$. For example, to explain $\Delta E (R_2)$ in the adiabatic-Fock basis $\{|g,n\rangle, |e,m\rangle\}$, it is straightforward to recognize that $|g,2\rangle$ couples with $|g,1\rangle$ through $\langle g,2|\hat{\boldsymbol\mu}(\hat{a}^{\dagger}+\hat{a})|g, 1\rangle=\boldsymbol\mu_{gg}\langle 2|(\hat{a}^{\dagger}+\hat{a})|1\rangle$, and $|g,1\rangle$ couples to the $|e,0\rangle$ through $\boldsymbol\mu_{ge}\langle 1|(\hat{a}^{\dagger}+\hat{a})|0\rangle$. Hence, the Rabi model that ignores the permanent dipole will not give a correct prediction. Under the usual Fock state basis, it is not conceptually intuitive to discuss the role of ${\boldsymbol\mu}_{gg}$ and ${\boldsymbol\mu}_{ee}$. Under the PFS basis, on the other hand, it is intuitive to understand these phenomena, and these coupling can be simply estimated as $V_\mathrm{IC}\langle n_\mathrm{C}|m_\mathrm{I}\rangle$, that means $\Delta E (R_2) = 2 V_\mathrm{IC}\langle 2_\mathrm{I}|0_\mathrm{C}\rangle$ (when ignoring other non-resonance couplings). In SI, we demonstrate that these simple analytic expressions of  $\langle n_\mathrm{C}|m_\mathrm{I}\rangle$ provides almost exact answer for $\Delta E (R_i)$ presented in this panel.

Fig.~\ref{down}c presents the time-dependent number of photons $\langle \hat{N}\rangle(t)=\langle\Psi(t)|\sum_{\alpha}\hat{b}^{\dagger}_{\alpha} \hat{b}_{\alpha} |\Psi(t)\rangle$ to demonstrate the downconversion process, where $|\Psi(t)\rangle$ is the total wavefunction of the hybrid molecule-cavity system. The initial condition is $|\Psi(0)\rangle \sim \exp(-\alpha(R-R_{g})^2)\otimes|\Phi_5(R)\rangle$, where $|\Phi_5(R)\rangle \approx |\mathrm{I},0_\mathrm{I}\rangle$ in the Franck-Condon region. The initial wavepacket is centered at $R_g=3.01$ a.u. and a width $\alpha = 19.12$ a.u. to mimic a vertical Franck-Condon excitation of the molecule-cavity hybrid system from its ground state. In the range of parameters used here, $\langle\hat{N}\rangle(t)$ reaches to as high as 2.4, representing multiple photons created per molecular excitation. The maximum value for $\langle\hat{N}\rangle (t)$ also increases with a higher $\chi$. 

Fig.~\ref{down}d presents the population of the polarized Fock states  $\hat{\rho}_{n}  =\mathrm{Tr}_{\alpha}[\sum_{\alpha}|\alpha, n_{\alpha}\rangle\langle n_{\alpha},\alpha|]$ at $\chi = 0.007$ a.u. With this coupling strength, all of the LIAC at $R_1$, $R_2$ and $R_3$ becomes considerably large. Hence, the wavepacket first branches at $R_3$, then at $R_2$ and finally at $R_1$, leading to  sequential rising  of the 3-photon (blue), 2-photon (red), and 1-photon population (green). This demonstrates the possibility of converting molecular excitation to multiple photons. It is also possible to selectively control the number of photons by changing $\chi$. More detailed discussions of the population dynamics is provided in SI, together with the results in the polariton basis $|\Phi_j(R)\rangle$. Note that the downconversion presented here is enabled due to the coupling between the $|\mathrm{I},n_\mathrm{I}\rangle$ (for $n\ge 2$) and $|\mathrm{C},0_\mathrm{C}\rangle$ states (and between $|g,n\rangle$ and the $|e,0\rangle$ state in the Fock state basis), which go beyond the prediction of the Rabi model.

\begin{figure}
 \centering
  \begin{minipage}[h]{1.0\linewidth}
     \centering
         \includegraphics[width=\linewidth]{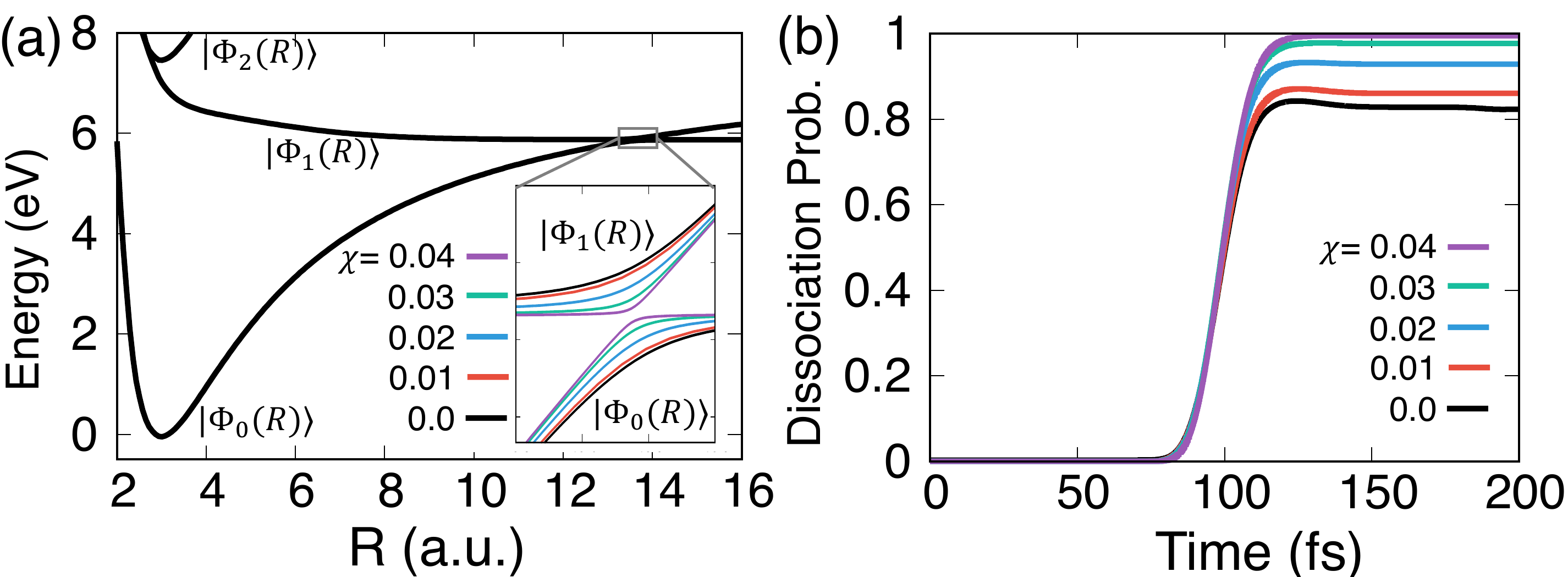}
       \end{minipage}%
   \caption{\small Controlling photochemical reactions with photon field polarization. (a) Polaritonic potentials with a high photon frequency $\hbar \omega_\mathrm{c} = 7.5$ eV, with the inset showing the lowest two polaritonic potentials near $R_\mathrm{0}$ at various $\chi$. (b) Dissociation probability at various $\chi$.}
 \label{Control}
\end{figure}

Fig.~\ref{Control} demonstrates that the electronic non-adiabatic coupling at $R_0$ can be modified through the non-orthogonality of the PFS to enhance the photo-dissociation dynamics. To clearly show this, we choose a high photon frequency $\hbar\omega_\mathrm{c} = $ 7.5 eV, such that all of the other polariton states are above $|\Phi_1(R)\rangle$ throughout the dynamically relevant parts of $R$. Fig.~\ref{Control}a presents the first three polaritonic potentials $E_j (R)$ of the hybrid system with the inset depicting the polariton potentials $E_0(R)$ and $E_1(R)$ at different $\chi$. The polaritonic potentials of the $|\Phi_0(R)\rangle$ and $|\Phi_1(R)\rangle$ states are nearly identical to the original molecular adiabatic potentials of $|g\rangle$ and $|e\rangle$ state. At $\chi = 0$, the energy-splitting between $|\Phi_0(R)\rangle$ and $|\Phi_1(R)\rangle$ corresponds to the bare molecular system, given by $2 V_\mathrm{IC}(R_0)$. By increasing $\chi$ we see a clear trend of decreasing the energy-splitting, as indicated in the inset. The splitting between the two states is given by $V_\mathrm{IC}(R) \langle  0_\mathrm{C}| 0_\mathrm{I}\rangle  =  V_\mathrm{IC}(R) e^{-{1\over2}[\chi\Delta\mu_\mathrm{IC}(R_{0})/{\hbar\omega_\mathrm{c}]^2}}$ (ignoring all other off-resonant contributions), where $\Delta\mu_\mathrm{IC} (R_0)= \mu_\mathrm{I}(R_0) - \mu_\mathrm{C}(R_0)$. Thus, increasing $\chi$ effectively decrease $\Delta E(R_0)$, causing the non-adiabatic coupling $\langle\Phi_0 (R)|\nabla_R|\Phi_1 (R)\rangle$ to increase (see SI). 

Fig.~\ref{Control}(b) presents the photo-dissociation dynamics of the LiF molecule defined as $\langle\Psi (t)|\Phi_0(R)\rangle\langle \Phi_0(R)|\Theta(R-R_0)|\Psi (t)\rangle$, where $\Theta$ is the heaviside function. The initial quantum state is $|\Psi(t=0)\rangle \sim \exp(-\alpha(R-R_g)^2)\otimes|\Phi_1 (R)\rangle $. The dissociation occurs by making a non-adiabatic transition from the initially occupied $|\Phi_1(R)\rangle$ state to the dissociative $|\Phi_0(R)\rangle$ state around the $R_0$ region. With an increasing $\chi$, due to the decreasing energy-splitting and increasing the non-adiabaticity causes a larger non-adiabatic transition probability. Therefore, enhanced dissociation dynamics occurs from increasing the light-matter coupling $\chi$. Despite several existing works on controlling chemical reactivity through molecule-cavity coupling~\cite{Triana2018,Triana2019, csehi2019ultrafast,Vendrell2018,Kowalewski2016,Feist2018} that relied on introducing new non-adiabatic couplings through resonance light-matter interactions to modify chemical reactivity, the control scheme demonstrated here is fundamentally different. Here, we modify the original electronic non-adiabatic coupling through an off-resonance light-matter interactions. Due to the choice of an off-resonant photon mode, no cavity photons are emitted to modify the chemical reactivity, in contrast to most of the previous works that involve emission and absorption of cavity photons. Thus, the cavity loss is expected to play a minimal role in the polariton photochemistry dynamics presented here. 


In conclusion, we demonstrated that the presence of the permanent dipole moments and the associated dipole self-energy terms leads to the polarization of the vacuum photon field.  These polarized Fock states associated with different electronic states are non-orthogonal to each other.  This non-orthogonality is similar to the dynamical Casimir effect ~\cite{Moore1970,Yablonovitch1989,Schwinger1992,Dodonov10,Juan2020} where the permanent dipole difference plays a similar role as the physical displacement of the cavity mirrors. Through numerically exact quantum dynamics simulations, we demonstrate the possibility to exploit this non-orthogonality to achieve multiple photon generation and enhancing the photo-dissociation of a molecule by coupling to a cavity.

More importantly, we demonstrate the conceptual and computational convenience of the polarized Fock states in molecular cavity QED, compared to the widely used vacuum's Fock states. We envision that the polarized Fock representation will provide a powerful theoretical framework for future polariton chemistry investigations.



\begin{acknowledgments}
\subsection{Acknowledgments}
This work was supported by the National Science Foundation ``Enabling Quantum Leap in Chemistry" program under the Grant number CHE-1836546. Computing resources were provided by the Center for Integrated Research Computing (CIRC) at the University of Rochester. A.M. appreciates the support from his Elon Huntington Hooker Fellowship. S.M.V. appreciates a generous support from the i-scholar program of the Department of Chemistry at the University of Rochester. P. H. acknowledge the support from his Cottrell Scholar award. A.M. appreciates stimulating discussions with Wanghuai Zhou and Marwa Farag. We appreciate valuable conversations with Prof. Peter Milonni.
\end{acknowledgments}

\end{document}